\documentclass[a4paper,prb,twocolumn,showpacs]{revtex4}
\usepackage{hyperref}
\usepackage[pdftex]{graphicx}
\usepackage{amsmath}
\usepackage{array}
\usepackage{bm}
\usepackage{fancyhdr}
\usepackage{makeidx}
\usepackage{subfigure}
\usepackage{epsfig}
\usepackage{amsfonts}
\usepackage{amssymb}
\usepackage{color}

\begin{document}

\title{Electrical control of spin dynamics in finite one-dimensional systems}
\author{A. Pertsova, M. Stamenova and S. Sanvito}
\affiliation{School of Physics and CRANN, Trinity College Dublin, Dublin 2, Ireland}
\date{\today}

\begin{abstract}
We investigate the possibility of the electrical control of spin transfer in monoatomic chains incorporating spin-impurities. 
Our theoretical framework is the mixed quantum-classical (Ehrenfest) description of the spin dynamics, in the spirit of 
the $s$-$d$-model, where the itinerant electrons are described by a tight-binding model while localized spins are 
treated classically. Our main focus is on the dynamical exchange interaction between two well-separated spins. This 
can be quantified by the transfer of excitations in the form of transverse spin oscillations. We systematically study the 
effect of an electrostatic gate bias $V_g$ on the interconnecting channel and we map out the long-range 
dynamical spin transfer as a function of $V_g$. We identify regions of $V_g$ giving rise to significant 
amplification of the spin transmission at low frequencies and relate this to the electronic structure of the channel. 
\end{abstract}
 
\maketitle

\section{Introduction}\label{intro}

The rapid development of the field of spintronics\cite{Awschbook} over the past two decades has uncovered exciting and novel 
phenomena related to the dynamics of the electronic spin in a wide variety of systems, ranging from bulk materials to spatially-confined 
structures \cite{spindyn}. Fueled by the ever-growing needs for speed, capacity and energy-efficiency in computing, the central objective 
in understanding and ultimately controlling the spin properties in the solid state has been constantly shifting towards the nano-scale. 
At these lengths and times the conventional methods for spin control, based on magnetic fields alone, are limited by scalability issues. 
Alternative approaches are thus sought and they typically involve electric fields of some form~\cite{Nadjib}.

One way to manipulate spins by purely electrical means relies on the spin-transfer torque mechanism~\cite{Slonchewski}. This approach 
uses spin-polarized currents to control the direction of the magnetization and has been realized in various nano-structured materials 
ranging from magnetic multilayers~\cite{storque} to, more recently, single atoms in STM-type geometries~\cite{stmspin}. Alternative to 
electric current control is the optical control, such as in the laser-driven ultrafast magnetization switching~\cite{Rasing,RasingReview}. In 
a somewhat different context, the optical manipulation of single spins in bulk media~\cite{sspin} is at the heart of the most promising 
candidates for the quantum information processing technology~\cite{diamond1, diamond2}.

Another alternative is based on the idea of an electrostatic control of the spin-density, i.e. of the construction of spin-transistor 
type devices\cite{Jansen}. Recently, the concept of gate-modulated spin-pumping~\cite{Tserkovnyak} transistors has been studied  
theoretically in infinite graphene strips with patterned magnetic implantations~\cite{Mauro_gate}. Importantly, such devices rely on the 
efficient transport of spin information between two points in space and time, and require the possibility to actively tune the propagating 
spin-signal during its transport. In this work we explore this possibility for atomistic spin-conductors. We consider a finite mono-atomic 
wire linking two localized spin-carrying impurities. When one of the localized spins is set into precession, it generates a perturbation 
in the spin-density. This perturbation is carried through the wire by conducting electrons and can be detected in the dynamical response 
of the second spin. We show that the propagation of the spin-signal and consequentially the dynamical communication between the two 
spin centers, can be tuned by means of an electrostatic gate applied to the interconnecting wire. Our main finding is an enhancement 
of the communication for a certain range of gate voltages. This is linked to the modification of the electronic structure of the wire induced
by the applied electrostatic gate.

Because of their considerable size the systems investigated here are still beyond the present numerical capabilities of first-principles 
dynamics~\cite{abinitio} and as such they are described by model Hamiltonians. Usually the dynamics is approached within the linear 
response approximation~\cite{Mauro_gate}. In this paper we go beyond the linear response limit and propose a fully microscopic 
description based on the time-dependent Schr\"{o}dinger equation, which allows us to describe arbitrary excitations. In other words our 
simulations are not limited to small gate voltages or small-angle spin precession. We use a single-band tight-binding Hamiltonian to model 
the itinerant $s$-electrons in our metallic wires and include local Heisenberg interactions to a number of magnetic ions (typically one or 
two). The latter are modeled as classical spin degrees of freedom and enter on equal footing in the common mixed quantum-classical 
dynamic portrait of the system. 

This paper is organized as follows. In Section \ref{theo} we introduce the model system and our  theoretical framework. Section \ref{res1} 
contains the results of our investigations. Firstly we investigate the electron response of an atomic wire, not including magnetic impurities, 
to local spin excitations. Secondly, we study the dynamical interplay between two spin-impurities electronically connected by the wire and 
show how such a dynamics is affected by the gate voltage. Finally we draw some conclusions.

\section{The model}\label{theo}

\begin{figure}[ht]
\begin{center}\includegraphics[scale=0.3,angle=270]{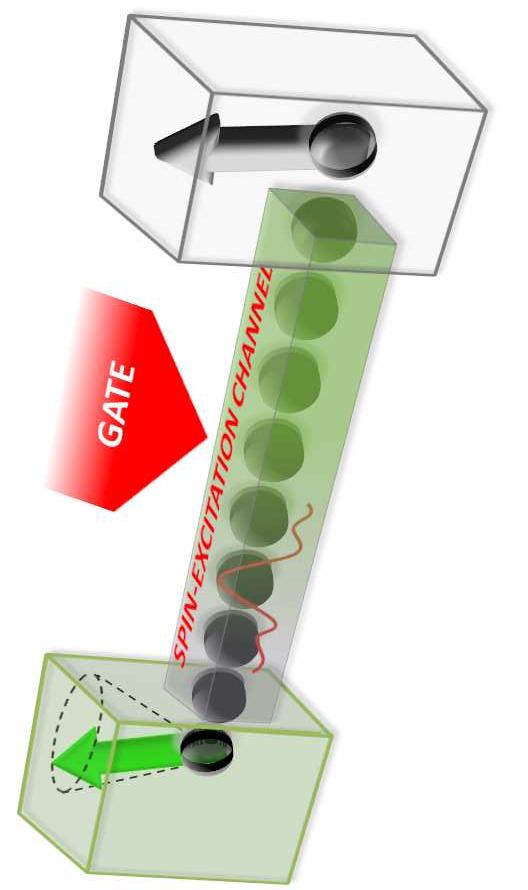}\end{center}
\caption{(Color online) Model system investigated in this work: two localized spins are electronically connected by a mono-atomic 
wire, where electrons can flow. An electrostatic gate is applied to some of the interconnecting sites. }\label{model}
\end{figure}

A cartoon of the device considered is presented in Fig. \ref{model}. This consists of an $N$-sites long atomic wire interconnecting two magnetic 
centers. The latter, represented in the figure by the spin-vectors, enter our model as substitutional magnetic impurities positioned at the two ends 
of the wire (the local spin originates from the deeply localized $d$-electrons). One of them, say the left-hand side spin-center $\boldsymbol{S}_{1}$, 
is labeled as ``driven'', as its precession is induced and sustained by a local (at that site) magnetic field. The other localized spin, $\boldsymbol{S}_{2}$, 
located at the other end of the chain is the ``probe'' spin and it is not directly coupled to any external magnetic field but only to the electron gas. 
In this way $\boldsymbol{S}_{2}$ probes the magnetic excitations produced by the driven spin $\boldsymbol{S}_{1}$ as these propagate through 
the interconnecting wire. The device is described as a mixed quantum-classical system in the spirit of the ($s$-$d$)-model~\cite{Kondo}, where the 
time-dependent Hamiltonians of the two exchange-coupled spin sub-systems read
\begin{eqnarray}
\hat{H}_\mathrm{el}(t)\!\! & = & \!\!\!\!\!\! \sum_{\scriptsize \begin{array}{cc} i,j=1,N \\ \alpha=1,2 \end{array}} \!\!\!\!\!\! \, H_{ij}^\mathrm{TB} \, c^{\alpha\dagger}_{i} \,c^{\alpha}_{j} \\
& & - J \!\! \sum_{\alpha, \beta=1}^{2} \left[ \boldsymbol{S}_{1}(t) c^{\alpha\dagger}_{1}\, c^{\beta}_{1} + \boldsymbol{S}_{2}(t) \, c^{\alpha\dagger}_{N}\, c^{\beta}_{N} \right]\cdot \hat{\boldsymbol{\sigma}}^{\alpha\beta}\:, \nonumber  \label{eq:1}
\end{eqnarray}
\begin{equation}
H_\mathrm{S}(t)=-\boldsymbol{S}_{1}(t)\cdot \left[ J \boldsymbol{s}_1(t) + g\mu_\mathrm{B} \boldsymbol{B}\right] - J \boldsymbol{S}_{2}(t)\cdot\boldsymbol{s}_N(t)\:. \label{eq:1b}
\end{equation}
The top expression is for the quantum electrons. Here $H_{ij}^\mathrm{TB}=\varepsilon_{i}\,\delta_{ij} +\gamma\,\delta_{i,i\pm 1}$ 
is a single-orbital tight-binding Hamiltonian with on-site energies $\varepsilon_{i}$ and hopping integral $\gamma$  ($\gamma$ sets the 
relevant  energy scale for the entire system);  $c_{i}^{\alpha\dagger}(c_i^{\alpha})$ is the creation (annihilation) operator for an electron 
with spin-up ($\alpha=1$) or spin-down ($\alpha=2$) at the atomic site $i$; $\boldsymbol{\hat{\sigma}}=\frac{1}{2}(\sigma_{x}, \sigma_{y},\sigma_{z})$ 
is the electron spin operator, $\{\sigma_{l}\}_{l=x,y,z}$ being the set of Pauli matrices; $\boldsymbol{S}_{1,2}$ is the unit vector in the direction 
of the localized spin;  $J\,>\,0$ is the exchange coupling strength. 

The classical dynamics of the local spins is governed by $H_\mathrm{S}(t)$, which describes the interaction of $\boldsymbol{S}_{1,2}$ 
with the mean-field local electron spin-density $\boldsymbol{s}_i \equiv \left\langle \boldsymbol{\hat{\sigma}}\right\rangle_i$, taken as the 
instantaneous expectation value of the conduction-electron spin at site $i$ (see further in text for the exact definition). The classical 
Hamiltonian also includes interaction with the external magnetic field $\boldsymbol{B}=(0,0,B_z)$, which is applied locally to $\boldsymbol{S}_1$ 
only, in order to drive its precession. We further assume $g=2$ for the localized spins and $\mu_\mathrm{B}=5.788\,10^{-5}$eV/T is the Bohr magneton.

In order to study the dynamics of this system in the time-domain we integrate the coupled quantum and classical equations of motion 
(EOM) of the two spin sub-systems~\cite{maria,maria2}. In our spin-analogue to Ehrenfest molecular dynamics\cite{ehrenfest}, the full set of 
coupled Liouville equations read 
\begin{eqnarray}
\dfrac{d\hat{\rho}}{dt} &=&\frac{i}{\hbar}\left[ \hat{\rho},\hat{H}_\mathrm{el}\right]  \label{eq:2}, \\
\dfrac{d\boldsymbol{S}_{n}} {dt} &=&\left\lbrace \boldsymbol{S}_{n},H_\mathrm{S}\right\rbrace = \frac{\boldsymbol{S}_n}{S} \times \left\{ \begin{array}{cc} J \boldsymbol{s}_1 + g \mu_\mathrm{B} \boldsymbol{B} \quad & \mathrm{for}\,\, n=1 \\  J \boldsymbol{s}_N \quad & \mathrm{for}\,\, n=2 \end{array} \right. \, . \nonumber
\end{eqnarray}
Here $\{,\}$ represents the classical Poisson bracket, $[,]$ stands for the quantum-mechanical commutator and $S=\hbar$ is the magnitude 
of the two classical spins. The first EOM is for the electron density matrix $\hat{\rho}$. At the initial time, $t_0$, $\hat{\rho}$ is constructed 
from the eigenstates $\{ |\varphi_{\nu}\rangle\}_{\nu=1}^{2N}$ of the spin-polarized electron Hamiltonian $\hat{H}_\mathrm{el}(t_0)$ 
[see Eq. (\ref{eq:1})], with the composite index $\nu=\{i,\sigma\}$ labeling the set of $2N$ spin-polarized eigenstates. We define 
$\rho(t_0)=\sum_{\nu} f_{\nu} |\varphi_{\nu} \rangle \langle \varphi_{\nu}|$ with  $f_{\nu}=\eta_F(\epsilon_{\nu}-E_F)$ being the occupation 
numbers distributed according to Fermi-Dirac statistics. The instantaneous onsite spin-density is thus generated as 
\begin{equation}
\boldsymbol{s}_i (t)\equiv \left\langle \boldsymbol{\sigma}\right\rangle_i (t) = \mathrm{Tr}\left[ \rho(t)  \boldsymbol {\sigma}\right]_{i}=\sum_{\alpha\beta} \rho_{ii}^{\alpha\beta}(t)\boldsymbol{\sigma}^{\beta\alpha}\:. 
\end{equation}
The coupled EOMs are integrated numerically by using the fourth-order Runge-Kutta (RK$4$) algorithm~\cite{compphys}. As a result the 
set of trajectories for the local electronic spin-densities $\boldsymbol{s}_{i}(t)$ are obtained as well as those of the driven and the probe 
classical spins $\boldsymbol{S}_{1,2}(t)$. The typical length of the chain that we consider is $N=100$.

The effect of an electrostatic gate is incorporated in our model as a rigid shift of the onsite energies 
$\varepsilon_i \rightarrow \varepsilon_i + V_g $, where $i\in[i_1,i_2]$ is a certain range of sites in the middle of the chain 
and $V_g$ is the gate voltage \footnote{This region can also be re-interpreted as part of a different material so that the same 
model system also describes a one-dimensional tri-layered heterostructure in which the two peripheral layers are identical.}.

\section{Dynamics of the itinerant spins with frozen impurities}\label{res1}

\subsection{No external gate}

As a preliminary step towards the combined quantum-classical dynamics we first address the dynamics of the spin-density 
of the itinerant electrons in the presence of the two local spins, whose directions are fixed, e.g. 
$\boldsymbol{S}_{1,2} \vert \vert \hat{\boldsymbol{z}}$. A spin excitation is produced by a small but finite spatially-localized 
perturbation in the spin-density. As initial density matrix at $t_0$ we use the one that corresponds to a perturbed Hamiltonian
$\hat{H}_\mathrm{el} (t_0-\delta t) $ in which one of the localized spins is slightly tilted in the $x-z$ plane, such 
that $\boldsymbol{S}_1 (t_0-\delta t) \cdot \hat{\boldsymbol{z}}\approx d\theta \lesssim 5^\mathrm{o}$. However at 
$t_0$ we bring $\boldsymbol{S}_1$ back to its original direction ($||\hat{\boldsymbol{z}}$) where it stays throughout the simulation.
In other words we study the time evolution of the system with the two local spins parallel to each other starting from the 
ground state electronic charge density of the system where one of the two spins is tilted by a small angle.

The evolution of the density matrix for $t>t_0$ is then given by 
$\hat{\rho}(t)=e^{-i\hat{H}_\mathrm{el}t/\hbar} \hat{\rho} (t_0) e^{i\hat{H}_\mathrm{el}t/\hbar}$, which translates into the following 
expression for the density matrix elements
\begin{equation}
\rho_{kl}^{\sigma\sigma\prime}(t)= \sum_{mn}e^{-i\omega_{mn}t}\,c_{km}^{\sigma}\, c_{ln}^{*\sigma\prime} \sum_{i\alpha}\sum_{j\beta} 
c_{im}^{*\alpha}\,c_{jn}^{\beta}\, \rho_{ij}^{\alpha\beta}(t_0) \,. \label{eqspec}
\end{equation}
In Eq.~(\ref{eqspec}) we have defined the frequencies $\omega_{mn}\equiv e_{m}-e_{n})/\hbar$ corresponding to differences 
between the eigenvalues $e_{m}$ of $\hat{H}_\mathrm{el}(t_0)$. The coefficients 
$c_{km}^{\sigma} = \left\langle k \sigma|\right. \left. \varphi_{m}\right\rangle$ are the projections of the eigenvectors $|\varphi_{m}\rangle$ 
on the spin-resolved atomic orbital basis $\left. |k\sigma\right\rangle\equiv\left.|k\right\rangle\otimes\left.|\sigma\right\rangle$, where 
$k$ represents the atomic site and $\sigma=\uparrow,\downarrow$ is the spin component. 

\begin{figure}[ht!]
\begin{center}\includegraphics[
  scale=0.29,angle=0,clip=true]{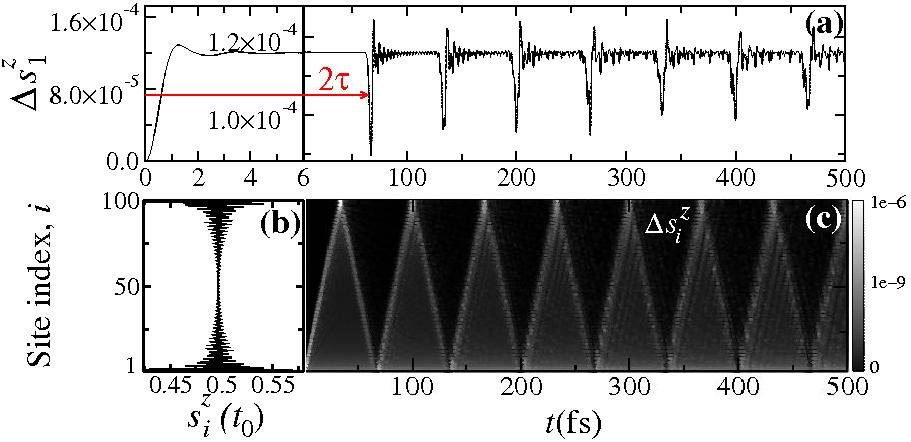}
\end{center}
\caption{(Color online) Spin excitation of one dimensional wire. (a) Time dependence of the spin-polarization on the first site 
($\Delta s^{z}_{1}$) as obtained from the numerical integration of the EOMs and directly from Eq. (\ref{eqspec}). (b) Initial 
spin-polarization $s^{z}(t_0)$ as a function of site index. (c) Time and space evolution of $\Delta s^{z}_{i}$ with the color shade 
representing the magnitude of absolute value of  $\Delta s^{z}_{i}(t)$. Note that $2\tau$ is the wave-packet round-trip time as 
also seen in the bottom panel.}\label{elspin}
\end{figure}

The typical evolution of the spin density of the itinerant electrons resulting from excitation described above is presented in 
Fig.~\ref{elspin}. The trajectories of the individual onsite spin polarizations $\Delta s^z_{i}(t)=s^z_{i}(t)-s^z_{i}(t_0)$ stemming 
from from Eq.~(\ref{eqspec}) are perfectly identical (on the scale of the graph) to those obtained by the numerical integration of the 
EOM [i.e. from Eq. (\ref{eq:1})], confirming the reliability of our time-integrator for the typical duration of the simulations. 

The spatial distribution of the initial spin polarization, $s^z_{i}(t_0)$, is shown in Fig.~\ref{elspin}(b) and its subsequent evolution, 
presented in Fig.~\ref{elspin}(c) can be qualitatively characterized as the propagation of a spatially-localized spin wave-packet. 
This travels along the wire with a practically uniform velocity very close to the Fermi level group velocity, as expected in view of the 
rather small overall local spin-polarization of the chain. The packet then gradually loses its sharpness as it disperses. However, the 
backbone of the packet is detectable even after a few tens of reflections at the ends of the wire. Such a feature demonstrates that this 
finite 1D model atomic system is a relatively efficient waveguide for spin wave-packets (of sub-femtosecond duration) at least over 
the investigated time-scales of a few picoseconds.

Eq. (\ref{eqspec}) can also be used  to calculate directly the spectra of particular spin observables. These match perfectly with those 
calculated by performing the discrete Fourier Transform \cite{numrec} (dFT) of the time-dependent spin-densities obtained over 
the finite duration of the dynamic simulation. Clearly, in the absence of the localized impurities (i.e. for a finite homogeneous tight-binding 
chain) $\omega_{mn}$ can be calculated exactly. The spectrum of $\Delta s^z_{i}$ has monotonously decreasing amplitudes from 
the lowest possible frequency $\omega_\mathrm{min}\approx \frac{3\pi^{2}|\gamma|}{(N+1)^{2}\hbar}$  to the maximum one 
$\omega_\mathrm{max} \approx \frac{4|\gamma|}{\hbar}$ (expressed in the limit of $N\rightarrow\infty$). This is also true for the case 
of interaction with the two frozen local impurities (e.g. for $J=\gamma$) as these have a minor effect on the electron Hamiltonian. 
For the parameters typically used in our simulations the corresponding maximum period 
$T_\mathrm{max}=\frac{2\pi}{\omega_{min}}\lesssim 1$~ps is within the total time of the simulation while the corresponding minimum 
period $T_\mathrm{min}=\frac{2\pi}{\omega_{max}}\approx 1$ $fs$ is much larger than the typical time step $\Delta t=0.01$~fs. 

\begin{figure}[ht]
\begin{center}\includegraphics[
  scale=0.41,angle=0,clip=true]{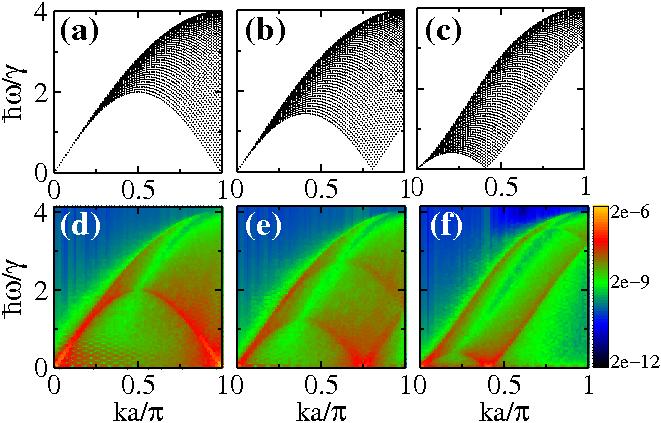}
\end{center}
\caption{(Color online) dFT$[\Delta s^{z}_{i}(t)](k,\omega)$ for different band-filling: (a, d) $\rho_0=0.5$ , (b, e) $\rho_0=0.6$ and 
(c, f) $\rho_0=0.8$. The top panels show the exact analytical excitation spectrum for a homogeneous tight-binding chain without local 
spins. The bottom panels show the results of our numerical simulations. The inter-site distance $a$ is arbitrary. The color shade in the 
bottom panel represents the absolute value of dFT$[\Delta s^{z}_{i}(t)](k,\omega)$ and is logarithmically scaled for better 
contrast.}\label{specfill}
\end{figure}

A more detailed spectral analysis is obtained by the two-dimensional dFT power portraits\cite{numrec} of  $\Delta s^z_{i}$ 
(see Fig. \ref{specfill}) denoted as dFT$[\Delta s^{z}_{i}(t)](k,\omega)$. In Fig. \ref{specfill} we compare such spectra for the itinerant 
spin-dynamics to its exact counterpart in the case of a finite  homogeneous tight-binding chain. These portraits reveal key features 
of the one-dimensional fermionic system that vary systematically with the band-filling $\rho_0$, namely (i) a near-continuum of 
allowed modes in a certain $(k,\omega)$-space region, defined by a low- and a high-energy dispersion functions, (ii) a linear 
dispersion for small $k$ ($\omega\propto k$ for $k\rightarrow 0$). An analogous mode-occupation patterns have been rigorously 
analyzed in relation to the dynamical properties of one-dimensional quantum Heisenberg spin chains which too have a tight-binding 
type dispersion relation~\cite{spin_chain}. The low-energy mode-occupation limit is due to the fact that the energy of the electron-hole 
excitation can approach zero only for $\Delta k\rightarrow 0$ and $\Delta k\rightarrow 2k_\mathrm{F}$, where $k_\mathrm{F}$ is the 
Fermi vector ($k_\mathrm{F}=\frac{\pi}{2a}$ for $\rho_0=0.5$, corresponding to one electron per site). The variation of the band filling 
away from the half-filling results into a folding of the low-energy limit as shown in Figs.~\ref{specfill}(b), \ref{specfill}(c), 
\ref{specfill}(e) and \ref{specfill}(f). Note that due to electron-hole symmetry we only show the spectra for $\rho_0 \geq 0.5$.

\subsection{Electrostatic gate applied} \label{res1_gate} 

\begin{figure}[ht]
\begin{center}\includegraphics[scale=0.35,clip=true,angle=0]{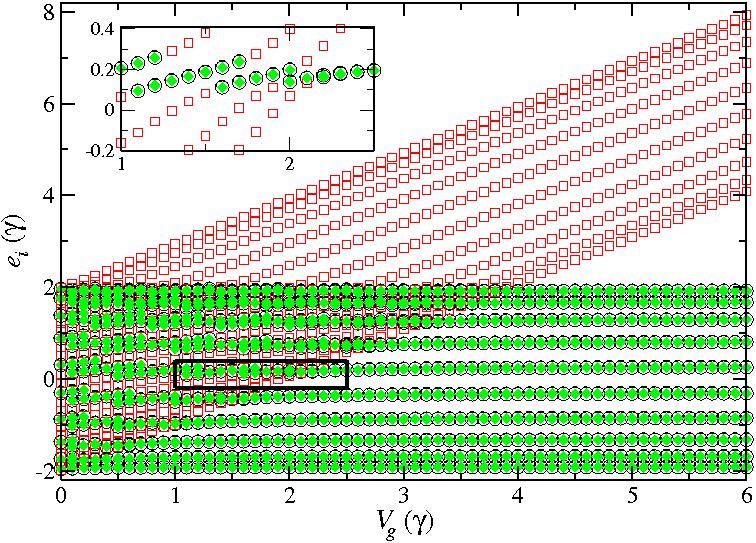}\end{center}
\caption{(Color online) Schematic of the gate-dependence of the ground state energy spectrum of $\hat{H}_\mathrm{el}$. According 
to their weights at the three subsections of the chain $\Omega_\mathcal{L,M,R} = \sum_{i\in \mathcal{L,M,R} } |c_{in}|^{2}$ we distinguish 
three types of states depending on which of the three partial weights is the greatest. We use black circles for the case of 
$\Omega_\mathcal{L}$, red squares for $\Omega_\mathcal{M}$ and green diamonds for $\Omega_\mathcal{R}$. Note that the 
eigenstates corresponding to the two ungated regions ($\Omega_\mathcal{L}$ and $\Omega_\mathcal{R}$) are degenerate by 
symmetry, as the gate is applied in the exact center of the wire. The inset shows a magnification of an area of intermediate $V_g$, 
illustrating the situation of avoided crossings. A short chain with $N=30$ sites is used for simplicity.} \label{eigengate}
\end{figure}

The dynamics of the itinerant spins, as described by Eq.~(\ref{eqspec}), in the case of an electrostatic gate applied to a section 
of the wire (typically in the middle of the wire) cannot be expressed in a closed form as a function of $V_g$ for arbitrarily big systems. 
As such we resort to our numerical integration scheme. Before addressing the dynamics, however, we first analyze the ground-state 
electronic structure of the gated wires for different values of the gate potential $V_g$. Displayed in Fig.~\ref {eigengate} is the adiabatic 
variation of the eigenvalues for a non spin-polarized chain with $N=30$ sites (without localized spins) as a function of the gate potential 
$V_g$ (the gate is applied to 10 sites in the middle of the chain, i.e. at the sites with indexes going from $i_1=11$ to $i_2=20$). Note that 
energy-related quantities on all the figures are in units of the hopping integral $\gamma$. For $V_g$ close to $0$, the discrete energy 
spectrum spans in the range $[-2\gamma, 2\gamma]$. With the increase of $V_g$ the levels spacing distorts as the eigenstates are affected 
differently by the gate. Certain eigenvalues grow nearly linearly with increasing $V_g$ (these are shown as red squares in 
Fig.~\ref{eigengate}). The spatial distribution of these eigenstates is predominantly concentrated in the gated region, i.e. they corresponds 
to the region where the local onsite energy has been modified.

For extremely large values of $V_g$ ($V_g > 4\gamma$)  the chain is effectively split into three energetically-decoupled parts, the 
gated middle and the two identical un-biased ends on the left-hand side and on the right-hand side. More interesting for us is the 
range of intermediate gate voltages, for which the three parts of the wire are substantially affected but not yet decoupled by the gate 
voltage. For such $V_g$ avoided crossings occur between states localized in the gated and non-gated regions. This gives rise to 
additional low-frequency lines in the dynamical spectrum as described by Eq.~(\ref{eqspec}). As we will demonstrate in the Appendix 
the presence of these avoided crossings around the Fermi level for certain intermediate $V_g$ yields an enhanced transmission through 
the gated wire (waveguide) at low frequencies.

\begin{figure}[ht]
\begin{center}\includegraphics[%
  scale=0.42,angle=0,clip=true]{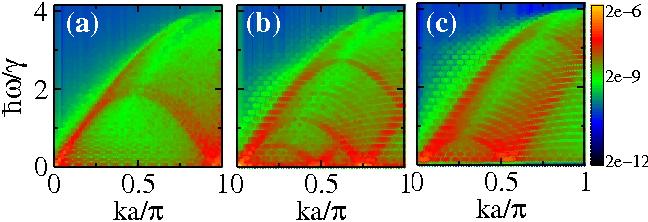}
\end{center}
\caption{(Color online) dFT$[\Delta s^{z}_{i}(t)](k,\omega)$ for different values of the gate potential: (a) $V_g=0.2\gamma$ , 
(b) $V_g=2.2\gamma$ and (c) $V_g=4.6\gamma$. The color shade is identical to that on Fig. \ref{specfill}.}\label{specgate}
\end{figure}

The two-dimensional dFT images (Fig. \ref{specgate}) of the spin dynamics in the presence of the gate bring additional dimension 
to the electron spectroscopy analysis. The difference here with respect to the case depicted in Fig.~\ref{specfill}(d) is that a gate has 
been applied to the middle of the chain in the ground state, i.e. $V_g \sum_{\alpha, i \in [i_1,i_2]} c^{\alpha\dagger}_i c_i$ has been 
added to $\hat{H}_\mathrm{el}$ at $t=t_0$. Again we use the half-filling case, $\rho_0=0.5$ (one electron per atom). From the figure it 
is immediately noticeable that the effect of the variation of $V_g$ on the excitation spectra is quite similar to the effect of the band-filling 
in the non-gated case. Due to the gate potential, the relative electron populations of the gated and gate-free parts of the chain change 
(the gated region is depopulated). For intermediate values of $V_g$ [see Fig. \ref{specgate}(b)] the excitation spectrum is rather a 
superposition of two spectra with band fillings below and above $0.5$ (approximately $0.2$ and $0.6$). Furthermore we find a 
substantially increased population of the low-frequency modes for all $k$-vectors, i.e. of states were forbidden by symmetry for $V_g = 0$. For large 
$V_g$  [see Fig. \ref{specgate}(c)] the middle part of the wire becomes almost completely depleted and the $(k,\omega)$-portrait 
corresponds effectively to the excitation spectrum of a single chain with a band-filling $\rho_0 \approx 0.75$, which is similar to the 
non-gated case presented in Fig. \ref{specfill}(f). 

\section{Combined quantum-classical spin dynamics in the presence of a gate}\label{res2}

\begin{figure}[ht]
\begin{center}\includegraphics[scale=0.3,clip=true,angle=0]{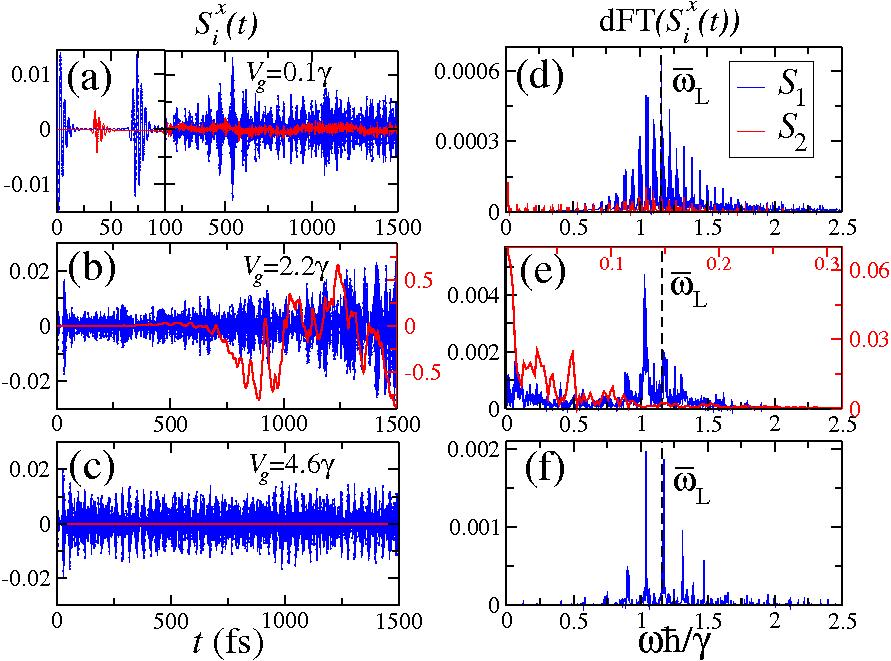}\end{center}
\caption{(Color online) Time evolution of the $x$-components of the localized spins, $S^{x}_{1}(t)$ (blue) and  $S^{x}_{2}(t)$ (red) 
[left] and the corresponding spectra, dFT$[S^{x}_{i}(t)](\omega)$, [right] for three different gate voltages (a, d) $V_g=0.1\gamma$, 
(b, e) $V_g=2.2\gamma$ and (c, f) $V_g=4.6\gamma$. The frequency-domain is represented by the dimensionless quantity 
$\hbar\omega/\gamma$ and $\bar{\omega}_{L}=\hbar\omega_{L}/\gamma=1.16$ for $\gamma=1$ $eV$. Note that in the case of 
$V_g=2.2\gamma$ different scales are used for the magnitudes of $S^{x}_{1}(t)$  and  $S^{x}_{2}(t)$ and their spectra (axes 
corresponding to  $S^{x}_{2}(t)$ are marked in red).}\label{gate1}
\end{figure}

The inclusion of dynamic local spin-impurities in the model requires the numerical integration of the set of coupled 
non-linear EOMs of Eq.~(\ref{eq:2}). For a small number of classical spins the dynamics of the itinerant spin-density is qualitatively 
very similar to what is described by Eq. (\ref{eqspec}). The frequency-domain analysis of the classical spin trajectories by means of 
dFT reveals the characteristic signature of the discrete electronic spectrum with only additional modulations in the amplitudes.
We focus now on the case in which $\boldsymbol{S}_1$ is driven by a local magnetic field into a precession, hence it acts as a 
spin-pump (see cartoon in Fig.~\ref{model}). It is important to note that what we refer to as a local magnetic field $\boldsymbol{B}=(0,0,B^z)$ 
is only instrumental to trigger and sustain a uniform Larmor precession of $\boldsymbol{S}_1$, i.e. it does not produce any Zeeman
splitting in the itinerant electrons spectrum. We consider now the following situation: the quantum-classical spin-system is in its ground 
state until $t=t_0$, when the first local spin $\boldsymbol{S}_1$ starts fluctuating to form a small misalignment with $\boldsymbol{B}$. 
This sets the entire quantum-classical spin system into motion. The typical trajectories of the transverse components 
${S}_1^x (t)$ and ${S}_2^x (t)$ and their frequency-domain dFT images for different values of the gate voltage are presented 
in Fig.~\ref{gate1}. 

In this case the excitation of a transverse spin-density at $t=0$ by the initiation of the Larmor precession is very similar in nature to 
the excitation induced by tilting one local spin investigated before (section \ref{res1}). In fact it indeed develops in a very similar way 
in the early stages of the time evolution. Just like in Fig.~\ref{elspin} a non-equilibrium transverse spin-density spin packet is sent along 
the wire and both the localized spins respond each time the packet reaches them. Figure~\ref{gate1}(a) displays the case of a very 
small gate voltage $V_g \ll \gamma$. The frequency-domain image of the precessional motion of $S^{x}_{1}(t)$ has the signature of 
the discrete electronic spectrum with an amplitude envelope that peaks at the Larmor frequency $\omega_\mathrm{L}=2\mu_B B_1/\hbar$. 
The dFT-portrait of the probe spin $S^{x}_{2}(t)$ is very similar to that of $S^{x}_{1}(t)$. Although significantly down-scaled in 
amplitude, it has the electronic-structure modes imprinted in its classical motion in the same way as it is for the driven spin. 
This qualitative behavior is observed for $V_g=0$ and a range of small voltages $V_g\le \gamma$ but it changes dramatically for 
gate voltages $V_g>\gamma$ [see Fig. \ref{gate1}(b) and (c)].  The case of extremely big voltages $V_g> 4\gamma$ (depicted in the 
bottom panel) is a trivial one for which the electrostatic barrier $V_g$ is completely opaque to the electrons and the probe spin 
$\boldsymbol S_2$ does not respond to the Larmor precession of $\boldsymbol S_1$ (note also the rarefaction of modes in the 
$S_1^x$ spectrum due to the effective shortening of the chain by 2/3). 

Special attention must be devoted to the intermediate $V_g$ regime in which the dynamics of the probe spin is qualitatively different 
and uncorrelated to that of the driven one [see Fig. \ref{gate1}(b)]. The typical outcome of the real time dynamics at such 
intermediate voltages $\gamma < V_g < 4 \gamma$ is that the probe spin starts accumulating very big transverse deflections. This is 
then fed back into the pumping spin which also deflects more but still preserves, to a great extent, its precession about the magnetic field. 

\begin{figure}[ht]
\centering
\begin{tabular}{cc} 
\epsfig{file=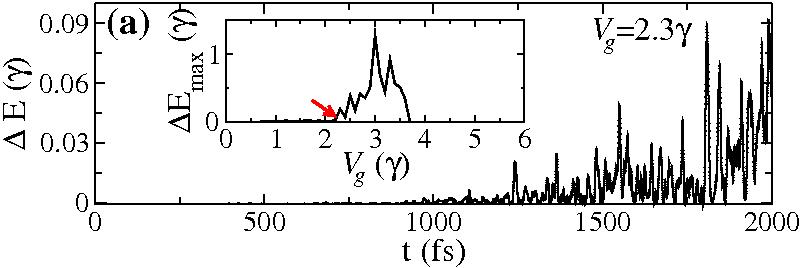,width=0.98\linewidth,clip=} \\
\epsfig{file=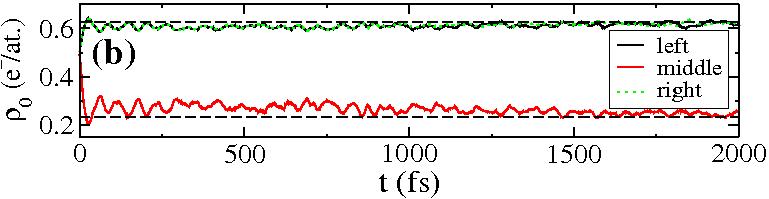,width=0.98\linewidth,clip=}
\end{tabular}
\caption{(Color online) (a) Time evolution of the energy transfer, $\Delta E(t)$, between electronic and localized spins for $V_g=2.2\gamma$. 
The inset shows the maximum value of the energy transfer, $\Delta E_{max}$, for a fixed duration of simulation (2\,ps) as a function of $V_g$. 
A red arrow indicates the particular voltage for which $\Delta E(t)$ is plotted. (b) Evolution of the electron populations of different parts of the chain 
for  $V_g=2.2\gamma$. The two black dashed lines show the equilibrium populations of the gate-free parts (upper line) and of the gated middle 
section (lower line), i.e. the populations corresponding to the case when the gate is applied in the ground state.}\label{energy}
\end{figure}

It is worth noting that this self-amplification of the spin-dynamics does not come at a total energy cost, as the model system described by 
Eg.~(\ref{eq:1}) is completely conservative. As such we simply observe a conversion of electrostatic energy into ``spin-energy'' in the form of 
a spin amplitude transverse to the driving magnetic field. This accumulation is related to the increasing energy transfer from the itinerant electrons 
to the localized moments, as illustrated in Fig.~\ref{energy}(a). Such an energy transfer is defined as 
$\Delta E(t)=|\Delta E_\mathrm{el}(t)-\Delta E_\mathrm{S}(t)|$, where $\Delta E_\mathrm{el/S}(t)= E_\mathrm{el/S}(t)-E_\mathrm{el/S}(t_0)$, 
$E_\mathrm{el}(t)=\mathrm{Tr}[\rho(t)\cdot H_\mathrm{el}(t)]$ and $E_\mathrm{S}(t)=-g\mu_{B}\boldsymbol{S}_{1}(t) \cdot \boldsymbol{B}$. 
The total energy conservation $\Delta E_{el}(t)=-\Delta E_S(t)$ has been verified within a relative error of $10^{-9}~\%$. The amplification of the 
energy transfer\footnote{This effect can appear as a particular form of  ``heating'' for the local spins, at the expense of electron gas, as within 
the time covered by the simulation ($\approx$10\,ps) the energy transfer seems to be irreversible. It is well-known, however, that the typical 
Ehrenfest molecular dynamics (EMD) suffers from the inability to account for Joule heating processes~\cite{Horsfield} due to the lack of dynamical 
correlations between the electrons and the ions. Our simulations are limited to picosecond times and by no means imply a possible qualitative 
difference of the EMD and its spin couterpart.} is characteristic of this $V_g$ range ($\gamma < V_g < 4 \gamma$) [see the inset of 
Fig.~\ref{energy}(a)].

The frequency-domain image of the localized spin trajectories [see Fig. \ref{gate1}(e)] for $\gamma < V_g < 4 \gamma$ shows a substantial 
qualitative difference for the driven and the probe spins. While around $\omega_\mathrm{L}$ the shape of the spectrum of $\boldsymbol{S}_1$ 
is similar to the two extreme $V_g$ cases, a significant amplitude is accumulated now at low frequencies. The main difference from the small 
$V_g$ case, however, is in the spectrum of $\boldsymbol S_2$ which shows practically no response at the Larmor frequency but has very large 
amplitudes in the low frequency range. 

This behaviour in the classical spin dynamics characteristic of certain gate voltages stems from the dominance of the low frequency modes in the 
electron dynamics at those $V_g$. The effect can be seen also in the much simpler case of a non-spin-polarized uniform atomic wire subjected to 
a sudden switch of an electrostatic gate in the middle. In the Appendix we consider the charge dynamics associated to this situation and analyze 
the lowest frequency components entering in Eq.~(\ref{eqspec}). The initial charge excitation, similarly to the transverse spin polarization in the model 
above, is triggered by a local electrostatic perturbation at the first site. It follows from Eq.~(\ref{eqspec}) that, for certain values of $V_g$ for which two 
adjacent adiabatic eigenenergies ($e_i,e_{i+1})$ around the Fermi level come closer to each other in an avoided crossing, the dynamical charge 
amplitude at the opposite end of the wire exhibits a peak at that particular very low frequency $(e_{i+1}-e_i)/\hbar$. In other words, a certain gate 
voltage condition is created for resonant charge transfer between the three parts of the chain at low frequencies. This, in the case of itinerant transverse 
spin-polarization, promotes the enhanced reaction of the probe spin to the spin pumping produced on the other side of the gate. 

\begin{figure}[ht]
\begin{center}\includegraphics[scale=0.48,angle=0,clip=true]{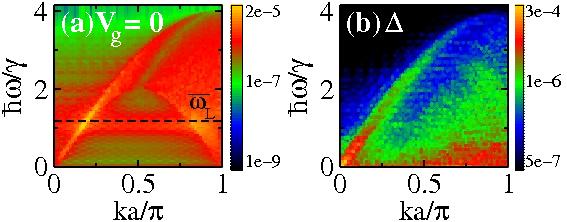}\end{center}
\caption{(a) dFT$[\Delta s^{z}_{i}(t)](k,\omega)$ for $V_{g}=0$. (b) The difference between dFT$[\Delta s^{z}_{i}(t)](k,\omega)$ for 
$V_{g}=2.2 \gamma$ and for $V_{g}=0$ defined as $\Delta=|\mathrm{dFT}[ s^{x}_{i}(t)](k,\omega)_{V_g=2.2\gamma}-\mathrm{dFT}[s^{x}_{i}(t)](k,\omega)_{V_g=0}|$. 
The color shade represents the absolute value of dFT$[\Delta s^{z}_{i}(t)](k,\omega)$ and $\Delta$, respectively, and is  logarithmically scaled 
for better contrast.}\label{specdiff}
\end{figure}

The electronic part of full quantum-classical spin dynamics in the presence of the gate (Fig.~\ref{gate1}) is, in many respects, similar to the charge-only case, 
described in Section~\ref{res1_gate} and in the Appendix. As before we construct the two-dimensional dFT portraits of the spin-density evolution. Providing 
additional $k$-resolved information these give another perspective for interpreting the dramatic increase of the low-frequency oscillations of the probe 
spin at certain gate voltages, when compared to the case of frozen $\boldsymbol{S}_i$ and electrons at $t=t_0-\delta t$ relaxed to the external gate potential 
(Fig.~\ref{specgate}). In the latter case, the increased low-frequency occupations (with respect to the non-gated case) were attributed to the split of the 
one-dimensional system into subsystems with different average electron densities. Here we find a similar dFT pattern, now in the presence of dynamically-coupled 
local spins and a gate abruptly introduced at $t=t_0$. Those two additional time-dependent factors result in a significant variation in time of the electron 
distributions in the gated and non-gated segments of the chain [see Fig. \ref{energy}(b)]. The dFT image in this case can be interpreted as a superposition 
of spectra corresponding to different $\rho_0$, resulting in a smearing of the low-frequency structure previously present in the $(k,\omega)$-space image. 
A comparison of the results for $V_g=0$  and $V_g=2.2\gamma$ is presented in Fig.~\ref{specdiff}. We find that the full dynamical $V_g=0$ portrait is rather 
similar to the results obtained for frozen-impurities (see Fig.~\ref{specgate}), the only additional feature being the adiabatic excitation at the Larmor frequency 
of the driven spin. The difference (point-by-point subtraction of the contour plots) with the corresponding gated case, depicted Fig.~\ref{specdiff}(b) shows 
that it is indeed mainly the lower frequency band of the spectrum (for any wave-vector) that gains occupation in the gated case.

\section{Conclusions}

In summary, we have employed atomistic dynamical simulations to investigate the effect of an electrostatic gate as means of controlling the indirect coupling 
between two distant localized spin impurities in a finite metallic wire comprising one hundred atoms. One of the impurities, precessing in external magnetic field, 
plays the role of a spin-pump and the response of the second (probe) spin was analyzed as a function of the gate potential applied to the interconnecting wire.
We identified a range of external gate potentials for which the spin-pumping is extremely efficient and leads to a substantial excitation of transversely-polarized 
itinerant spin density in the non-gated parts of the chain.  This, in turn, produces a massive low-frequency swing of the probe spin. Such a resonant effect has been 
related to gate-induced avoided crossings in the electronic structure of the interconnect. For certain gate voltages these occur in the vicinity of the Fermi level 
and assist the enhanced transfer of charge/spin across the barrier. Evidences for this effect are also identified in the Fourier portraits of the calculated time-dependent 
spin-distribution. 

\begin{acknowledgments}

This work is sponsored by Science Foundation of Ireland (Grant No. 07/IN.1/I945). Computational resources have been provided  by the Trinity Center for High 
Performance Computing and by the Irish Center for High-End Computing.
\end{acknowledgments}

\appendix*
\section{}

We consider the eigenstate-resolved dynamical amplitude $A_{mn}$ corresponding to a mode with frequency $\omega_{mn}$, associated
to the transitions between the eigenstates $e_m$ and $e_n$. The system is the simplified wire used for the spectroscopic analysis of Fig.~\ref{eigengate}, i.e. 
it is a non-spin-polarized 30-atom long chain. Analogously to the spin-polarized case, the $t=t_0$ excitation is applied simply as a potential shift at the 
first atomic site. From Eq.~(\ref{eqspec}) the dynamically-driven charge-accumulation amplitude at site $i$ is defined as
\begin{equation} 
A_{mn}^{i}=c_{im} c^*_{in} \sum_{kl}c^*_{km}c_{ln} \rho_{kl}(t_0) \,,   \label{Amni}
\end{equation}
such that
\begin{equation}
\rho_{ii}(t)=\sum_{mn}A_{mn}^{i} e^{-i\omega_{mn}t}\,
\end{equation}
is the dynamic electron occupation at site $i$. This quantity is calculated at the last site, $i=N$, i.e. at the opposite side 
of the chain with respect to the perturbation. In Eq.~(\ref{Amni}) $c_{im}=\langle i|\varphi_m\rangle$ are the onsite components 
of the eigenvectors $|\varphi_m\rangle$ of the non-spin-polarized  Hamiltonian $\hat{H}_\mathrm{el} (V_g)$ for which the gate 
voltage $V_g$ is applied (between the sites $i=11$ and $i=21$). 

\begin{figure}[ht]
\begin{center}
\includegraphics[width=0.98\linewidth,clip=true]{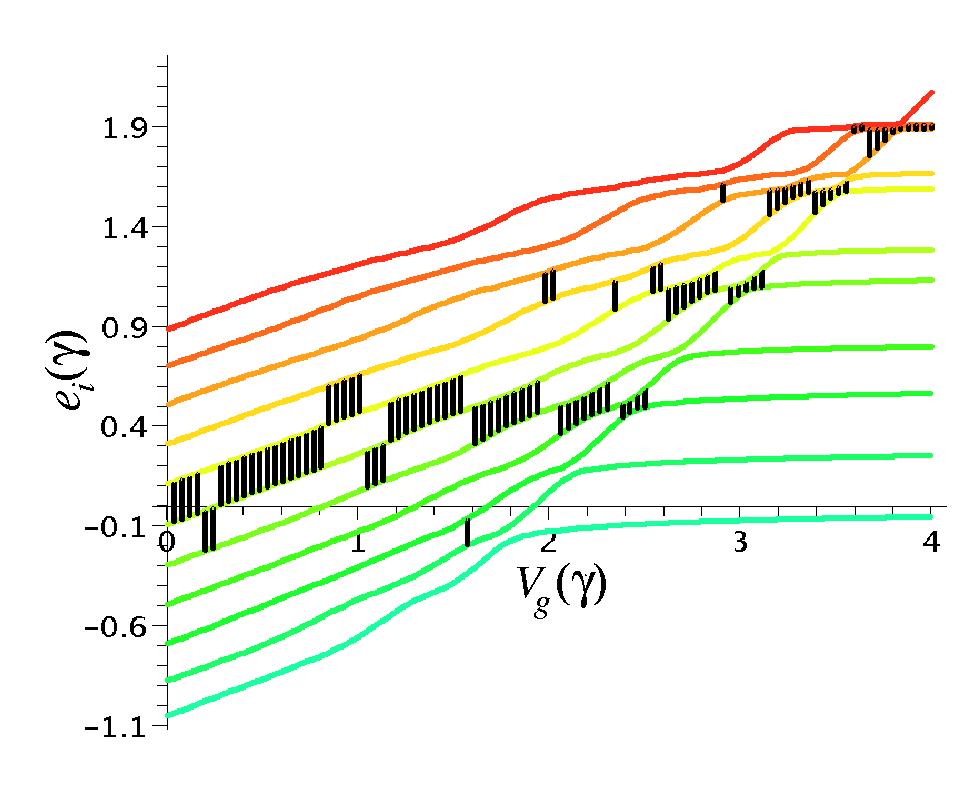}
\end{center}
\caption{(Color online) Adiabatic variation with the gate voltage of a few eigenvalues around the Fermi level (we have used 
$\varepsilon_0 =0$ and $|\gamma|=1$\,eV, hence $E_\mathrm{F}=0$ at $V_g=0$). Marked with black lines are the pairs of 
eigenstates $(i,i+1)$ which give rise to the highest (in absolute value) dynamical amplitude $A_{i,i+1}^{N}$ at the last site $N$.}\label{eigAmpl}
\end{figure}

From all the pairs of adjacent eigenstates $(e_i,e_{i+1})$, which give rise to the lowest frequency modes ($\omega_{i,i+1}=(e_{i+1}-e_i)/\hbar$) 
in the dynamics, we identify in Fig.~\ref{eigAmpl} those which have the highest (in absolute value) amplitude $A_{i,i+1}^{N}$. For small 
gate voltages these are the eigenstates around the Fermi level but as $V_g$ increases the modes with the largest amplitudes tend to arise 
where avoided crossings of eigenstates occur. These are also the modes with frequencies close to the lowest possible frequency for that
particular gate voltage.

\begin{figure}[ht]
\begin{center}
\includegraphics[width=0.98\linewidth,clip=true]{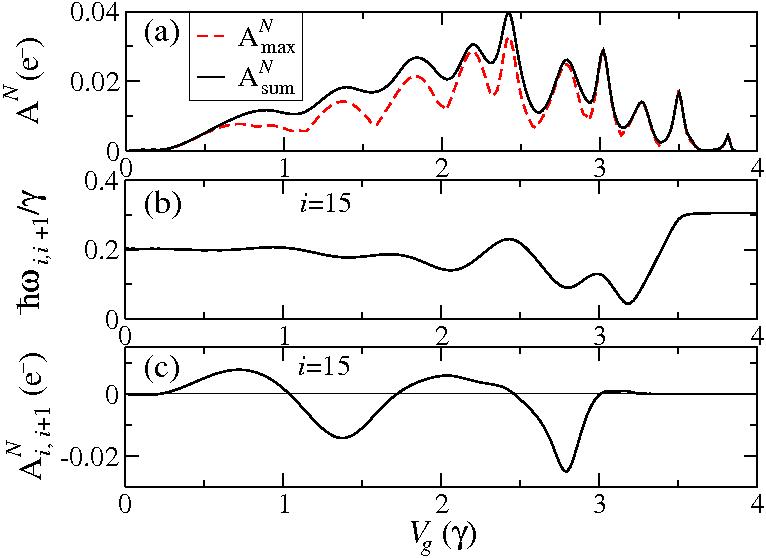}
\end{center}
\caption{(Color online) (a) The maximum single-mode amplitude $A_\mathrm{max}^N$ and an estimate for the total low-frequency amplitude 
$A_\mathrm{sum}^N$ (see text for details) as a function of the applied gate voltage. (b) The frequency of the mode related to the pair of 
eigenstates ($i=15$) around the Fermi level, adiabatically evolved with the gate voltage and (c) its corresponding amplitude as defined 
in Eq.~(\ref{Amni}).}\label{Ampls}
\end{figure}

The actual highest amplitude at the last site, corresponding to the highlighted modes in Fig. \ref{eigAmpl}, are depicted as a function of $V_g$ in 
Fig.~\ref{Ampls}(a), together with the quantity $A_\mathrm{sum}^N \equiv \sqrt{\sum_i (A_{i,i+1}^N)^2}$. This represents the pessimistic estimate (considering 
all modes orthogonal in phase) of the total low-frequency amplitude corresponding to all pairs of adjacent eigenvalues. Such curves demonstrate the resonant 
nature of the dynamical site-occupation as a function of the gate voltage. Fig.~\ref{Ampls}(a) also shows that the low frequency charge excitation peaks for 
intermediate gate voltages (for this case $V_g\sim2.5$\,$\gamma$) can be related to the observed enhanced low-frequency coupling between the itinerant and 
localized spins for the intermediate $V_g$ range. Interestingly, the two $A(V_g)$ curves coincide for very low and very high voltages, showing that in this case 
the entire dynamics is practically due to just one mode (marked in black in Fig. \ref{eigAmpl} ). In the intermediate regime it is clear that there are more than one 
low-frequency modes which have a significant amplitude, although the one marked in Fig.~\ref{eigAmpl} gives the leading contribution. 

For a particular pair of adjacent eigenstates there are a few instances where the adiabatic eigenvalues come closer together as a function of the gate voltage. 
In Fig.~\ref{Ampls}(b,c) we analyze the dependence of the frequency and the amplitude corresponding to the pair of eigenvalues just below and above the Fermi 
level in the ground state ($i=15$ for the 30-atom chain at half-filling). We find that the dynamical charge amplitude on the last site peaks (as absolute value) every 
time the frequency passes through a minimum.


\begin{thebibliography}{99}
\bibitem{Awschbook} T. Dietl, D. D. Awschalom, M. Kaminska and H. Ohno, \textit{Spintronics} (Elsevier, Amsterdam, 2008), Vol. 82.
\bibitem{spindyn} B. Hildebrand and K. Ounadjela, \textit{Spin dynamics in confined magnetic structures} (Springer-Verlag, Berlin, 2002).
\bibitem{Nadjib} N. Baadji, M. Piacenza, T. Tugsuz, F. D. Sala, G. Maruccio  and  S. Sanvito, Nat. Mat. \textbf{8}, 813 (2009).
\bibitem{Slonchewski}J. Slonczewski, J. Magn. Magn. Mater. \textbf{159}, L1 (1996).
\bibitem{storque} E. B. Mayers, D. C. Ralph, J. A. Katine,  R. N. Louie,  R. A. Buhrman, Science \textbf{285}, 867 (1999).
\bibitem{stmspin} S. Loth, K. von Bergmann, M. Ternes, A. F. Otte, C. P. Lutz and A. J. Heinrich, Nature Physics \textbf{6}, 340 (2010).
\bibitem{Rasing} A. V. Kimel, A. Kirilyuk, P. A. Usachev, R. V. Pisarev, A. M. Balbashov and Th. Rasing, Nature \textbf{435}, 655 (2005).
\bibitem{RasingReview} A.~Kirilyuk, A.V.~Kimel and Th.~Rasing, Rev. Mod. Phys. {\bf 82}, 2731 (2010).
\bibitem{sspin} R. Hanson and  D. D. Awschalom, Nature \textbf{453}, 1043 (2008).
\bibitem{diamond1} D. D. Awschalom, R. Epstein, and R. Hanson, Sci. Am. \textbf{297}, 84 (2007). 
\bibitem{diamond2} D. M. Toyli, C. D. Weis, G. D. Fuchs, T. Schenkel, and D. D. Awschalom, Nano Lett. \textbf{10}, 3168 (2010).
\bibitem{Jansen} R. Jansen, J. Phys. D: Appl. Phys. \textbf{36}, R289 (2003).
\bibitem{Tserkovnyak}Y. Tserkovnyak, A. Brataas, G. E. W. Bauer, Phys. Rev. Lett. \textbf{88}, 117601 (2002).
\bibitem{Mauro_gate} F. S. M. Guimarães, D. F. Kirwan, A. T. Costa, R. B. Muniz,  D. L. Mills, and M. S. Ferreira,  Phys. Rev. B \textbf{81}, 153408 (2010); F. S. M. Guimaraes, A. T. Costa, R. B. Muniz and M. S. Ferreira, Phys. Rev. B \textbf{81}, 233402 (2010).
\bibitem{abinitio} E. Runge and E. K. U. Gross, Phys. Rev. Lett., \textbf{52}, 997 (1984); K. L. Liu and S. H. Vosko, Can. J. Phys. \textbf{67}, 1015 (1989); V. P. Antropov, M. I. Katsnelson, M. van Schilfgaarde, and B. N. Harmon, Phys. Rev. Lett. \textbf{75}, 729 (1995); K. Capelle, G. Vignale and B. L. Gyrffy, Phys. Rev. Lett. \textbf{87}, 206403 (2001).
\bibitem{Kondo}J. Kondo, in \textit{Solid State Physics: Advances in Research and Applications}, edited by F. Seitz, D. Turnbull, and H. Ehrenreich (Academic, New York, 1969), Vol. 23, p. 184.
\bibitem{maria} M. Stamenova and S. Sanvito, in \textit{The Oxford Handbook on Nanoscience and Technology} (Oxford University Press, Oxford, 2010), Vol. 1, p. 169.
\bibitem{maria2} M. Stamenova, T.N. Todorov and S. Sanvito,  Phys. Rev. B \textbf{77}, 054439 (2008).
\bibitem{ehrenfest} A. P. Horsfield, D. R. Bowler and A. J. Fisher, J. Phys.: Condens. Matter \textbf{16}, L65 (2004).
\bibitem{compphys} J. Thijssen, \textit{Computational physics} (Cambridge University Press, 2007), p. 473.
\bibitem{numrec} W. H. Press,  S. A. Teukolsky, W. T. Vetterling and B. P. Flannery, \textit{Numerical recipes: the art of scientific computing} (Cambridge University Press, 2007), p. 600.
\bibitem{phys1D}T. Giamarchi, \textit{Quantum Physics in One Dimension} (Clarendon Press, Oxford, 2004), p. 12.
\bibitem{spin_chain}J. des Cloizeaux and J. J. Pearson, Phys. Rev. \textbf{128}, 2131 (1962); G. Muller and H. Thomas, Phys. Rev. B  \textbf{24}, 1429 (1981); O. Derzhko, in \textit{Condensed Matter Physics in the Prime of the 21st Century: Phenomena, Materials, Ideas, Methods}, proceedings of the 43rd Karpacz Winter School of Theoretical Physics, Ladek Zdroj, Poland, edited by J. Jedrzejewski (World Scientific, 2008), p. 35.
\bibitem{Horsfield} A. P. Horsfield, D. R. Bowler, A. J. Fisher, T. N. Todorov and M. J. Montgomery, J. Phys.: Cond. Mat. \textbf{16}, 3609 (2004) 
\end{thebibliography}
\end{document}